# The knowledge paradox: why knowing more is knowing less

Bruno Burlando

Dipartimento di Sienze e Innovazione Tecnologica
Università del Piemonte Orientale
Viale Teresa Michel 11
15121 Alessandria
Italy
e-mail : bruno.burlando@uniupo.it

**Abstract**

To provide an explanation of the evolution of scientific knowledge, I start from the assumption that knowledge is based on concepts, and propose that each concept about reality is affected by vagueness. This entails a paradox, which I term Knowledge Paradox (KP): i.e. we need concepts to acquire knowledge about the real world but each concept is a step away from reality. The KP provides a unifying context for the sorites and the liar paradoxes. Any concept is viewed as a sorites, i.e. it is impossible to set a boundary between what is, and what is not, the entity to which the concept refers. Hence, any statement about reality can be reduced to a liar, wherefrom the KP follows in its most general form: "If I know, then I do not know". The KP is self-referential but not contradictory, as it can be referred to two levels of knowledge: "if I know$_{epistemic}$, then I do not know$_{ontic}$", where the ontic level is made unachievable by concept vagueness. Such an interpretation of scientific knowledge provides an understanding of its dynamics. Concept proliferation within theories produces periods of knowledge decay that are episodically reversed by the formulation of new theories based on a smaller number of synthetic concepts.

.

**Introduction**

This article concerns an analysis of the meaning and progress of scientific knowledge. I will not make attempts at defining the limits of science, i.e. dealing with the long debated problem of demarcation (Resnik, 2000). Scientific knowledge will be simply defined here as a particularly structured form of conceptual knowledge, deriving from perceptual or instrumental experience of the world.

It is generally assumed that the purpose of science is to develop laws and theories in order to explain, understand and predict phenomena (Chalmers, 1990). Accordingly, it is believed that the scientific method has been one of the driving forces of human progress, while our knowledge about the world is usually related to the ability of theories to make predictions about reality (Gauch, 2003). However, progressive accumulation of scientific evidence can reveal shortcomings in extant paradigms, ultimately leading to theoretical rearrangements. Such a trend has been emphasized by Kuhn, with his distinction between normal and revolutionary phases in the progress of science (Kuhn 1962).

**Cognitive limits of concepts**

I start from the assumption that the failure of scientific theories derives from intrinsic cognitive limits of concepts. The term 'concept' will be used here in a wide acception, i.e. as an idea or notion that serves to designate a category of entities, or more in general, to provide a model of reality. Even though non-conceptual content of thought has been considered in the analysis of human cognitive processes and awareness (Bermudez, 1995), it is generally believed that conceptual content of thought is typical of scientific reasoning (Achinstein, 1968). Hence, our mind uses concepts in order to describe the real world, and such a process is generally considered an essential step in the advancement of science.

I wish to stress that the shortcomings of scientific paradigms, which may lead to a collapse of theoretical buildings, would depend on the reduction of the world to a set of concepts, i.e. discrete model objects, thereby yielding an artifactual vision of reality. A vivid reminiscence of these ideas can be found in the words of the French poet and philosopher Paul Valery: «Tout ce qui est simple est faux, mais tout ce qui ne l'est pas est inutilisable» (Guégan and Renaud, 2004). On a similar ground, my analysis holds that concepts can be viewed as an unavoidable simplification of reality, which inevitably leads to its falsification. These arguments are to some extent related to instrumentalism (White, 1943), which assumes that concepts and theories are merely useful instruments, or 'fictions', whose function is not to depict reality but rather to provide descriptions and predictions about the physical world. However, I am pushing my analysis further on, by proposing that concepts introduce elements of inconsistency within scientific theories that sooner or later lead to thorough revisions.

**Concept vagueness and the sorites**

The failure of concepts to provide a reliable description of the world addresses the problem of vagueness. This latter is strictly related to the sorites paradox, which entails the possibility that the borderline between a predicate and its negation be fuzzy (Tye, 1994; Keefe, 2000). In his analysis of the sorites, James Cargile refers to the impossibility of setting an exact point of transition between a developing tadpole and an adult frog (Cargile, 1969). It may be assumed that a proper example of vagueness had to be chosen, whereas in contrast my argument is aimed at showing that any real entity would have been appropriate, since I assume that vagueness is intrinsic to concepts and, therefore, any concept is susceptible to the sorites.

Obviously, the vagueness of concepts does not affect the practical use of the real objects to which they refer. In general, vagueness is ruled out whenever a certain approximation of reality can be tolerated in order to accomplish a particular project, as commonly occurs in everyday life, technical fields and professional work. Conversely, the vagueness of concepts heavily affects the rigorous in-depth analyses that need to be deployed in scientific activities.

Fundamental concepts of physics, like matter, time, and space, providing basic reference frames to scientific theories, are affected by vagueness. After the arising of quantum mechanics, there has been growing awareness that space and time are vague on a microscopic scale (Dolan, 2005). Moreover, their vagueness does not disappear on a macroscopic scale, as shown by the problem of infinite divisibility highlighted by Zeno's paradoxes (Mueller 1969), while their unwarranted use can lead to the puzzling situations of time travel paradoxes (Nahin, 1999). Also, the concept of matter has been made uncertain by quantum physics, under which the identity of elementary particles is suggested to be irreducibly vague (French and Krause, 2003; Chibeni, 2004).

In the field of biology, the formulation of the concept of cell has been a milestone of enormous theoretical importance (Baker, 1953). However, the identity of cells within organisms remains vague, due to the occurrence of syncytial tissues, endocytobiosis, incomplete cell divisions, and exchange of material between a cell and its surroundings. Similar reasoning can be applied to a number of biological entities.

In summary, science is affected by vagueness at its deepest foundations. The formulation of so-called exact laws is only possible if an appropriate armamentarium of concepts allows the edification of theoretical systems within which these laws are meaningful, as occurs e.g. for gravitation, electromagnetism and thermodynamics. By contrast, the lack of a suitable set of concepts renders extremely difficult the determination of exact laws for nonlinear systems, such as fluid turbulence, living beings, and stock markets (Bar-Yam, 1997).

**Knowledge Paradox and the progress of science**

Assuming that any concept about reality generates a sorites, since it involves some discretization of the real world, it follows that the formulation of concepts gives rise to vagueness and move us away from reality. This configures a paradox that I call the Knowledge Paradox (KP): scientific activities yield discoveries that are implemented within extant theories by means of new, ad hoc concepts, thus producing increasingly complex conceptual frameworks that progressively put our vision of reality out of focus. Consequently, the accumulation of concepts within theories does not produce an advancement of knowledge, but rather a progressive decay.

In order to clarify such a counterintuitive idea, I will use an example from the field of molecular biology. If a new, bona-fide protein is found, say protein A, a new concept is needed to implement such a discovery within the molecular theory of life. However, it is not clear if the notion of protein A refers to the protein's native form, to one of its post-translational modifications, or to one of the interactions established by the protein with other molecules, including the bulk of solvent water molecules. Now, given that each molecule of protein A is in a particular chemico-physical status as regards to its interactions with other molecules, a specific concept would be needed for each protein A molecule. Yet, even if this impracticable conceptualization could be done, it would not solve the problem, as molecular interactions are subjected to continuous change along time. Hence, the concept of the protein A is vague, representing at most an approximation of reality, and consequently it adds vagueness to the molecular theory of life. As stated above, this does not imply that the new concept cannot be helpful at all, but its use will be only profitable within theoretical boundaries entailing a simplification of reality. Conversely, if the aim is to achieve a better understanding of reality, the new concept will eventually become a hindrance.

As shown above, the accumulation of concepts negatively affects the description of reality. However, from time to time exhaustive revisions of data may succeed in producing synthetic theories that explain facts by a smaller number of new, revolutionary concepts. The evolution of scientific knowledge would therefore consist of episodic advancements, characterized by a reduction in the number of concepts, interspersed among long periods of decay characterized by concept accumulation (Fig. 1). Such a pattern can be easily recognized. By introducing his theory of natural selection, Darwin (1859) provided a synthetic explanation for the astonishing set of life diversity that had been collected through the years. About one century later, the theory of the genetic code yielded a unifying conceptual basis for biological heredity (Crick, 1967). Wegener's continental drift (1915), and the ensuing theory of plate tectonics, provided a synthetic explanation for the movement of Earth's plates, earthquakes, volcanoes, oceanic trenches, and mountain range formation. Also, by introducing the concept of "quark" in the 1960s, Gell-Mann and Zweig offered a unifying account of the variety of hadron particles described during the first half of the 20$^{th}$ century (Lichtenberg and Rosen, 1980). Similar features are shared by other major scientific breakthroughs, such as those carried out by Newton, Maxwell and Freud (Kuhn, 1962).

### Knowledge Paradox and the liar

If no theory can describe the real world in a completely affordable way, this must hold true for the KP itself. In other words, if any concept is vague when applied to reality, then the concept of 'concept' is vague too, and consequently also the KP is affected by vagueness. Hence, the KP makes a statement that, in the ultimate analysis, concerns its own meaning, and therefore is equivalent to the liar paradox (Martin, 1978). The liar is endowed with self-referentiality, or in other terms, it implies and is implied by its negation (Suber, 1990). Hence, it represents a contradiction built into the structure of language, while on the other hand the KP entails an endless loop within the structure of knowledge. Similar to the antinomy of the liar: "If this sentence is true, then it is not true", the KP can be formulated as: "If I know, then I do not know", which seems to make contradiction unavoidable.

A solution to the liar has been proposed by Tarski (1956) who, in his approach to the semantic concept of truth, adopted a hierarchy of languages consisting of object-language, to which the definition of truth is applied, and of meta-language, providing the definition of truth for object-language. By following a similar approach, knowledge can be analyzed from the standpoint of a meta-knowledge level, by making a distinction between our ideas about physical entities, viz. the epistemic level, and the real status of these entities, viz. the ontic level (Atmanspacher & Primas, 2003). Such a distinction allows a Tarskian reformulation of the KP providing two meanings for 'I know', whereby it should read: "If I know$_{epistemic}$, then I do not know$_{ontic}$".

According to this view, the formulation of concepts generates a gap between ontic and epistemic knowledge that cannot be amended, since any attempt at reducing this gap can only be carried out by formulating other concepts, and so on. Hence, the KP sets the limits of knowledge, while its self-referentiality highlights the unamendability of these limits.

A symptom of the ontic/epistemic disjunction of knowledge is the arising of inconsistencies in scientific theories that prelude the advent of their crisis. In Physics, striking examples have been the discoveries of the photoelectric effect and the black body radiation, which raised inconsistencies in classic physics and opened the way to quantum mechanics (Planck, 1901). In Biology, the Central Dogma of molecular genetics holds that biological information flows from DNA or RNA to proteins, and not the reverse (Crick, 1958). However, the discovery of prions (Prusiner, 1982), and the finding in yeast that these proteins can transmit information through generations and modify the expression of the genome (True & Lindquist, 2000), has posed a serious challenge to the overall consistency of the Dogma.

**Inconsistency of knowledge**

Combined analyses of the liar and the sorites have been carried out with the aim of finding a unifying treatment (Tappenden, 1993). The KP provides a link between the sorites and the liar by showing that any statement about the world is, or is entailed by, a statement that makes use of a sorites, and can therefore be reconducted to a liar. In the above example of protein A, the sorites emerges from the impossibility of setting a boundary between what is protein A and what is not protein A. Hence, the liar follows: "If this entity is (epistemologically) protein A, then it is not (ontologically) protein A".

In summary, knowledge is intrinsically inconsistent. So far as we formulate new concepts, then, the more concepts we have, the more inconsistencies we raise, the more we worsen knowledge about the world. By contrast, if we manage to reduce the number of concepts, by formulating synthetic ones, we reduce inconsistencies and eventually obtain an improvement of knowledge. So, contrarily to what it may be intuitively assumed: the more concepts, the less knowledge, and vice versa.

**Figures**

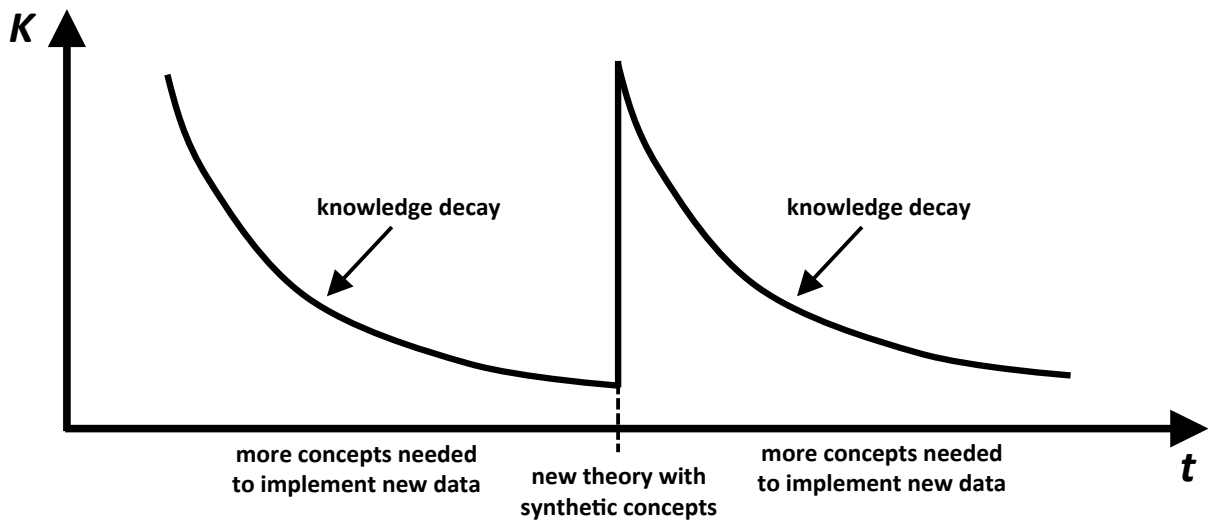

Figure 1. A hypothetical knowledge index (*K*), varying along time (*t*), is defined as being inversely proportional to the number of concepts used within a theory. So far as increasing numbers of concepts are needed to implement newly acquired data, the index scales down, indicating a decay of knowledge. However, from time to time a thorough revision of data gives rise to a synthetic theory consisting of a few innovative concepts. This renders many concepts obsolete, thus producing a sudden improvement of scientific knowledge.